\documentclass[]{nature}
\makeatletter\if@twocolumn\PassOptionsToPackage{switch}{lineno}\else\fi\makeatother

\usepackage{tabulary,graphicx,amsmath,amsfonts,amssymb}
\usepackage[utf8x]{inputenc}
\usepackage[T1]{fontenc}

\makeatletter
\renewenvironment{figure}
               {\@float{figure}}
               {\end@float}
\renewenvironment{figure*}
               {\@dblfloat{figure}}
               {\end@dblfloat}

\renewenvironment{table*}
               {\@dblfloat{table}}
               {\end@dblfloat}

\makeatother

\usepackage{url,multirow,morefloats,floatflt,cancel,tfrupee}
\makeatletter

\AtBeginDocument{\@ifpackageloaded{textcomp}{}{\usepackage{textcomp}}}
\makeatother
\usepackage{colortbl}
\usepackage{xcolor}
\usepackage{pifont}
\usepackage[nointegrals]{wasysym}
\makeatletter

\def\mcWidth#1{\csname TY@F#1\endcsname+\tabcolsep}

\def\cAlignHack{\rightskip\@flushglue\leftskip\@flushglue\parindent\z@\parfillskip\z@skip}
\def\rAlignHack{\rightskip\z@skip\leftskip\@flushglue \parindent\z@\parfillskip\z@skip}

\@ifundefined{etal}{}{}

\usepackage{ifxetex}
\ifxetex\else\if@twocolumn\@ifpackageloaded{stfloats}{}{\usepackage{dblfloatfix}}\fi\fi

\AtBeginDocument{
\expandafter\ifx\csname eqalign\endcsname\relax
\def\eqalign#1{\null\vcenter{\def\\{\cr}\openup\jot\m@th
  \ialign{\strut$\displaystyle{##}$\hfil&$\displaystyle{{}##}$\hfil
      \crcr#1\crcr}}\,}
\fi
}

\AtBeginDocument{%
  \@ifpackageloaded{endfloat}%
   {\renewcommand\efloat@iwrite[1]{\immediate\expandafter\protected@write\csname efloat@post#1\endcsname{}}}{\newif\ifefloat@tables}%
}%

\def\BreakURLText#1{\@tfor\brk@tempa:=#1\do{\brk@tempa\hskip0pt}}
\let\lt=<
\let\gt=>
\def\processVert{\ifmmode|\else\textbar\fi}

\@ifundefined{subparagraph}{
\def\subparagraph{\@startsection{paragraph}{5}{2\parindent}{0ex plus 0.1ex minus 0.1ex}%
{0ex}{\normalfont\small\itshape}}%
}{}

\newcommand\role[1]{\unskip}
\newcommand\aucollab[1]{\unskip}
  
\@ifundefined{tsGraphicsScaleX}{\gdef\tsGraphicsScaleX{1}}{}
\@ifundefined{tsGraphicsScaleY}{\gdef\tsGraphicsScaleY{.9}}{}
\def\checkGraphicsWidth{\ifdim\Gin@nat@width>\linewidth
	\tsGraphicsScaleX\linewidth\else\Gin@nat@width\fi}

\def\checkGraphicsHeight{\ifdim\Gin@nat@height>.9\textheight
	\tsGraphicsScaleY\textheight\else\Gin@nat@height\fi}

\def\fixFloatSize#1{}
\let\ts@includegraphics\includegraphics

\def\inlinegraphic[#1]#2{{\edef\@tempa{#1}\edef\baseline@shift{\ifx\@tempa\@empty0\else#1\fi}\edef\tempZ{\the\numexpr(\numexpr(\baseline@shift*\f@size/100))}\protect\raisebox{\tempZ pt}{\ts@includegraphics{#2}}}}

\AtBeginDocument{\def\includegraphics{\@ifnextchar[{\ts@includegraphics}{\ts@includegraphics[width=\checkGraphicsWidth,height=\checkGraphicsHeight,keepaspectratio]}}}

\DeclareMathAlphabet{\mathpzc}{OT1}{pzc}{m}{it}

\def\URL#1#2{\@ifundefined{href}{#2}{\href{#1}{#2}}}

\def\UrlOrds{\do\*\do\-\do\~\do\'\do\"\do\-}%
\g@addto@macro{\UrlBreaks}{\UrlOrds}

\edef\fntEncoding{\f@encoding}

\makeatother

\newif\ifmultipleabstract\multipleabstractfalse%
\usepackage[nomarkers,tablesfirst,nolists]{endfloat}

\def\fixFloatSize#1{}

\begin{document}

\title{Gemini and Physical World: Large Language Models Can Estimate the Intensity of Earthquake Shaking from Multi-Modal Social Media Posts}

\author{S.Mostafa Mousavi$^{1}$\thanks{Corresponding author.}\hspace{.4pc}\thanks{E-mail: mousavim@google.com}~,
        Marc Stogaitis$^{1}$,
        Tajinder Gadh$^{1}$,
        Richard M Allen$^{12}$,
        Alexei Barski$^{1}$,
        Robert Bosch$^{1}$,
        Patrick Robertson$^{3}$,
        Nivetha Thiruverahan$^{1}$,
        Youngmin Cho$^{1}$,
        Aman Raj$^{1}$ }

\maketitle 
\begin{affiliations}
  \item 
    Google LLC; Mountain View, CA, USA
  \item 
    Seismological Laboratory, University of California, Berkeley; Berkeley, CA, USA
  \item 
    Google Germany GmbH, Munich, Germany\end{affiliations}

\section*{\large{Abstract}}
\begin{abstract}
This paper presents a novel approach to extract scientifically valuable information about Earth's physical phenomena from unconventional sources, such as multi-modal social media posts. Employing a state-of-the-art large language model (LLM), Gemini 1.5 Pro\cite{reid2024gemini}, we estimate earthquake ground shaking intensity from these unstructured posts. The model's output, in the form of Modified Mercalli Intensity (MMI) values, aligns well with independent observational data. Furthermore, our results suggest that LLMs, trained on vast internet data, may have developed a unique understanding of physical phenomena. Specifically, Google's Gemini models demonstrate a simplified understanding of the general relationship between earthquake magnitude, distance, and MMI intensity, accurately describing observational data even though it's not identical to established models. These findings raise intriguing questions about the extent to which Gemini's training has led to a broader understanding of the physical world and its phenomena. The ability of Generative AI models like Gemini to generate results consistent with established scientific knowledge highlights their potential to augment our understanding of complex physical phenomena like earthquakes. The flexible and effective approach proposed in this study holds immense potential for enriching our understanding of the impact of physical phenomena and improving resilience during natural disasters. This research is a significant step toward harnessing the power of social media and AI for natural disaster mitigation, opening new avenues for understanding the emerging capabilities of Generative AI and LLMs for scientific applications.

\end{abstract}\def\keywordstitle{Generative AI, Large Language Models, Earthquakes, Natural Hazards}


\section*{Introduction}
Earthquakes pose significant risks to life and property, with the potential to cause widespread devastation. Due to their unpredictable nature and potential for catastrophic consequences, mitigating the devastating impact of earthquakes requires a comprehensive approach. This encompasses preparedness measures, early warning systems, and effective post-earthquake response strategies). Earthquake early warning (EEW) systems play a crucial role by providing timely alerts before the arrival of strong ground shaking, enabling individuals to take protective actions such as "drop, cover, and hold on"\cite{allen2019earthquake}. Furthermore, effective post-earthquake responses include search and rescue operations, damage assessment, and infrastructure restoration, which are critical to saving lives\cite{zhan2022post}. By combining preparedness measures, early warning systems, and rapid response strategies, communities can significantly reduce the impact of earthquakes and promote resilience in earthquake-prone regions.

The Modified Mercalli Intensity (MMI) scale offers a qualitative measure of earthquake shaking intensity based on observed effects, ranging from imperceptible shaking to catastrophic destruction, designated by Roman numerals I to XII\cite{wood1931modified}. Unlike magnitude scales that quantify the energy released at the source of the earthquake, the MMI scale focuses on the severity of shaking at specific locations, considering impacts on people, structures, and the environment. This is crucial as shaking intensity can vary significantly for earthquakes of the same magnitude \cite{minson2021shaking} due to factors like fault movement type and shallow geological properties. Moreover, the earthquake impact may vary depending on construction standards in different regions. By considering these factors, the MMI scale provides valuable insights into the actual experiences and damage observed at a given site, making it particularly useful for communicating earthquake impacts to the public\cite{musson2000intensity}.

In the realm of EEW systems, the MMI scale is used to estimate the appropriate level of alert and guide immediate safety measures\cite{allen2012intensity}.  In post-earthquake responses, MMI guides emergency responders in prioritizing areas for assistance and assessing the extent of damage\cite{worden2016shakemap, wald2005shakemap, wald1999utilization}. In earthquake engineering, MMI data is used to improve building codes and design structures that can withstand specific levels of ground shaking. It also finds utility in other domains, including insurance and disaster planning. Despite its wide-ranging applications, there exist several practical challenges associated with measuring MMI. One primary challenge stems from the subjective nature of MMI, as it is contingent upon observed effects rather than instrumental measurements. Another challenge lies in inconsistencies in reporting and/or the significant variability of MMI over short distances, attributable to factors such as complex local geological structures. Lastly, measuring MMI in real time poses difficulties, as it necessitates the involvement of trained observers for data collection. These challenges can introduce uncertainties into MMI estimations.

The advent of social media platforms has fundamentally transformed the way information is disseminated during natural disasters and crises. Social media users are now recognized as valuable providers of timely information, enabling the characterization of physical-world events\cite{hughes2009twitter}. They often share real-time updates, eyewitness accounts, and multimedia content, offering a rich source of data for disaster response and research purposes\cite{earle2010omg}. In the context of earthquake disaster management, several studies have shown the application of social media data for earthquake detection and damage area identification\cite{earle2011twitter, sakaki2012tweet, flores2017lightweight, bossu2018lastquake}, as well as for MMI estimation\cite{cresci2014towards, burks2014rapid, d2016real, mendoza2019nowcasting}. Furthermore, the increasing prevalence of CCTV cameras in urban and rural areas provides an unprecedented amount of real-world CCTV footage and video postings from social media platforms that can be harnessed for scientific research. This footage offers a unique opportunity to study natural hazards such as earthquakes, floods, and wildfires in unprecedented detail. By analyzing CCTV footage, scientists can track the spatiotemporal evolution of these events, identify previously hidden patterns and relationships, and gain deeper insights into the underlying physical processes. Additionally, it can be used to validate and calibrate numerical models, which are essential tools for predicting the behavior of natural hazards and assessing their potential impacts. However, extracting useful knowledge from social media data presents significant challenges, including issues of misinformation, irrelevant content, and language variations. Sophisticated method developments are required to overcome these challenges and extract valuable insights from the often noisy data\cite{imran2015processing}.

Generative Artificial Intelligence (GenAI) and Large Language Models (LLMs) have experienced remarkable advancements in recent years, demonstrating successful applications in various domains. These powerful tools hold immense potential for further enhancing the capabilities of crowdsourcing in earthquake studies and risk mitigation efforts. GenAI and LLMs possess the ability to extract pertinent information from unstructured social media posts. Moreover, they can analyze this information to estimate the intensity of local ground shaking with unprecedented ease and flexibility. This capability paves the way for novel approaches to understand collective behavior patterns\cite{zhou2012social}, gain insight into the dynamics of information propagation during crisis situations\cite{palen2016crisis, bagrow2011collective}, rapidly evaluate the impact of earthquakes, guiding emergency response efforts, provide situational awareness\cite{yin2012using}, and automatically collect valuable macroseismic data from unconventional sources. By combining the collective wisdom of social media users with AI-driven analytics, stakeholders such as emergency responders, policymakers, and researchers can gain a comprehensive understanding of disaster events. This includes real-time insights into the affected areas, and the severity of shaking. Consequently, EEW systems can be improved, and resilience strategies can be enhanced based on this real-time, user-generated information.
    
\section*{Results}
\textbf{Data}

YouTube, X, and TikTok (each ~33 percents) serve as the primary sources of data collected and analyzed in experiments of this study. Although official Application Programming Interfaces (APIs) are available for automated data retrieval, at this stage, we opted to collect data in the form of screenshots and screen recordings of relevant posts containing text, images, audio, and/or videos. This approach allows us to directly evaluate the ability of LLMs to extract and analyze relevant information from unstructured data in its final form as presented to humans. By doing so, we avoid the additional challenges and complexities associated with working with each platform's specific API, such as rate limits, access restrictions, and data formatting inconsistencies for our explorative work. We collected our dataset by searching for earthquake-specific keywords, including terms like "earthquake," "tremor," "shaking," and specific earthquake event names. We then removed posts curated by professionals or those whose authenticity or associated earthquake could not be confidently identified. Recognizing the crucial role of location in studying spatially variable ground shaking intensity, we limited our data to posts that included at least one location identifier. These identifiers could be mentions of the post/observation location in the text or speech, or visual cues in the provided image or video, such as recognizable landmarks or prominent distinctive buildings.

This procedure resulted in a total of approximately 84 posts documenting observed or experienced ground shaking caused by 7 earthquakes occurring in the United States (North California, South California, Oklahoma, and New Jersey), Japan (Noto and Tohoku), and Taiwan (Hualien) between March 2011 and mid-April 2024, spanning a magnitude range of 4.6 to 9.1 Mw (Table \ref{event_table}). The collected data can be categorized into two main groups: a) CCTV footage capturing the moment of shaking, and b) social media posts or news interviews where individuals share their personal experiences of feeling the earthquake. It is important to acknowledge the potential limitations of this data collection methodology. The focus on specific social media platforms and keyword-based search may introduce bias into the dataset. To address this, we employed multiple verification methods, including cross-referencing with official earthquake reports and news articles, to ensure the authenticity of the collected data and the associated earthquake events

\begin{figure}[ht!]
    \centering
    \includegraphics[width=0.95\columnwidth]{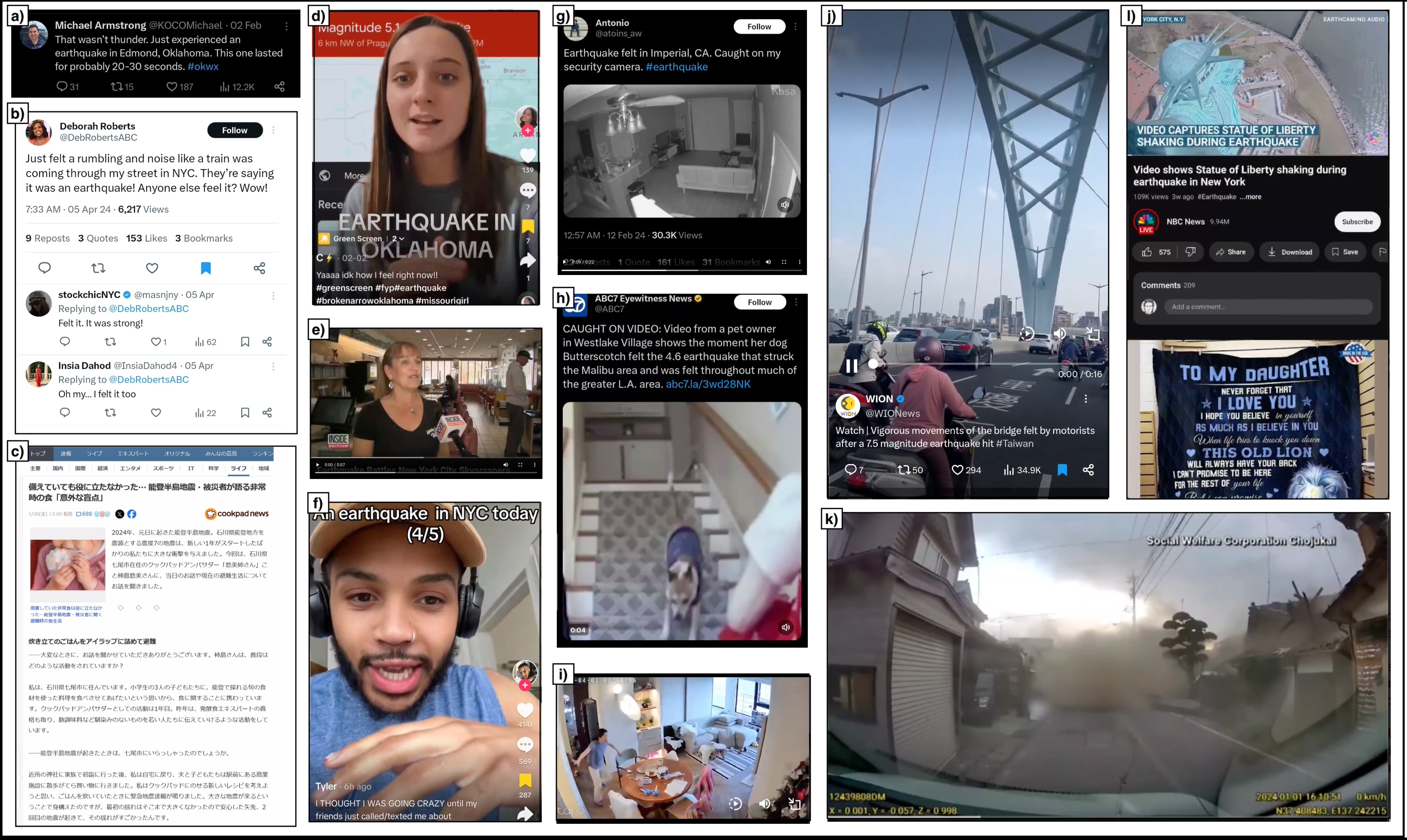}
    \caption{showcases the diversity of the dataset used in this study, which comprises screenshots and screen recordings of social media posts documenting individual (e.g., a, d, e and f) or group (e.g., b) experiences of earthquake shaking, as well as the responses of animals (e.g., h). These posts encompass a range of formats, including images (b) and videos (d to k) containing textual information, presented in various languages (e.g. c), sizes/durations, and background settings. The video content spans both indoor (g, h, and i) and outdoor (j, k, and l) environments. Indoor videos primarily consist of CCTV footage capturing the moment of earthquake shaking, while outdoor videos include similar CCTV footage as well as recordings of infrastructure damage. Additionally, the dataset incorporates post-earthquake narrative videos where individuals describe their personal experiences and observations during the earthquake (e.g. d to f).}
    \label{fig:dataset}
\end{figure}

To assess the model's ability to attend to and extract relevant information (textual, visual, and/or auditory) from diverse social media posts, we intentionally included screenshots and recordings with varying characteristics. These variations encompassed different video lengths, cropping sizes, languages, and background colors, simulating the wide range of formats encountered in real-world social media content. The captured views ranged from those containing the opinion or experience of a single individual, such as a short tweet, to those encompassing a tweet and a selection of its replies or comments. These replies and comments could be relevant, confirming the experience of the same earthquake or providing supplementary information about the earthquake source, or they could be irrelevant, such as jokes, unrelated discussions, or even misinformation. Additionally, both video and image views might contain unrelated information, such as advertisements or background visuals, intentionally introduced to assess the model's ability to focus on the earthquake-related content (Figure \ref{fig:dataset}).

The content within the posts provides valuable insights into the subjective experiences of individuals, including descriptions of shaking intensity, emotional responses, and reports of damage to personal belongings or surrounding structures. The inclusion of irrelevant information is crucial for evaluating the model's performance in real-world scenarios, where social media posts often contain extraneous content that is not pertinent to the earthquake event. By testing the model's ability to discern relevant information from noise, we can gain a better understanding of its effectiveness and robustness in extracting valuable insights from the complex and dynamic landscape of social media data.

\textbf{Comparison of Gemini's Output With Independent Data}

\begin{figure}[ht!]
    \centering
    \includegraphics[width=0.95\columnwidth]{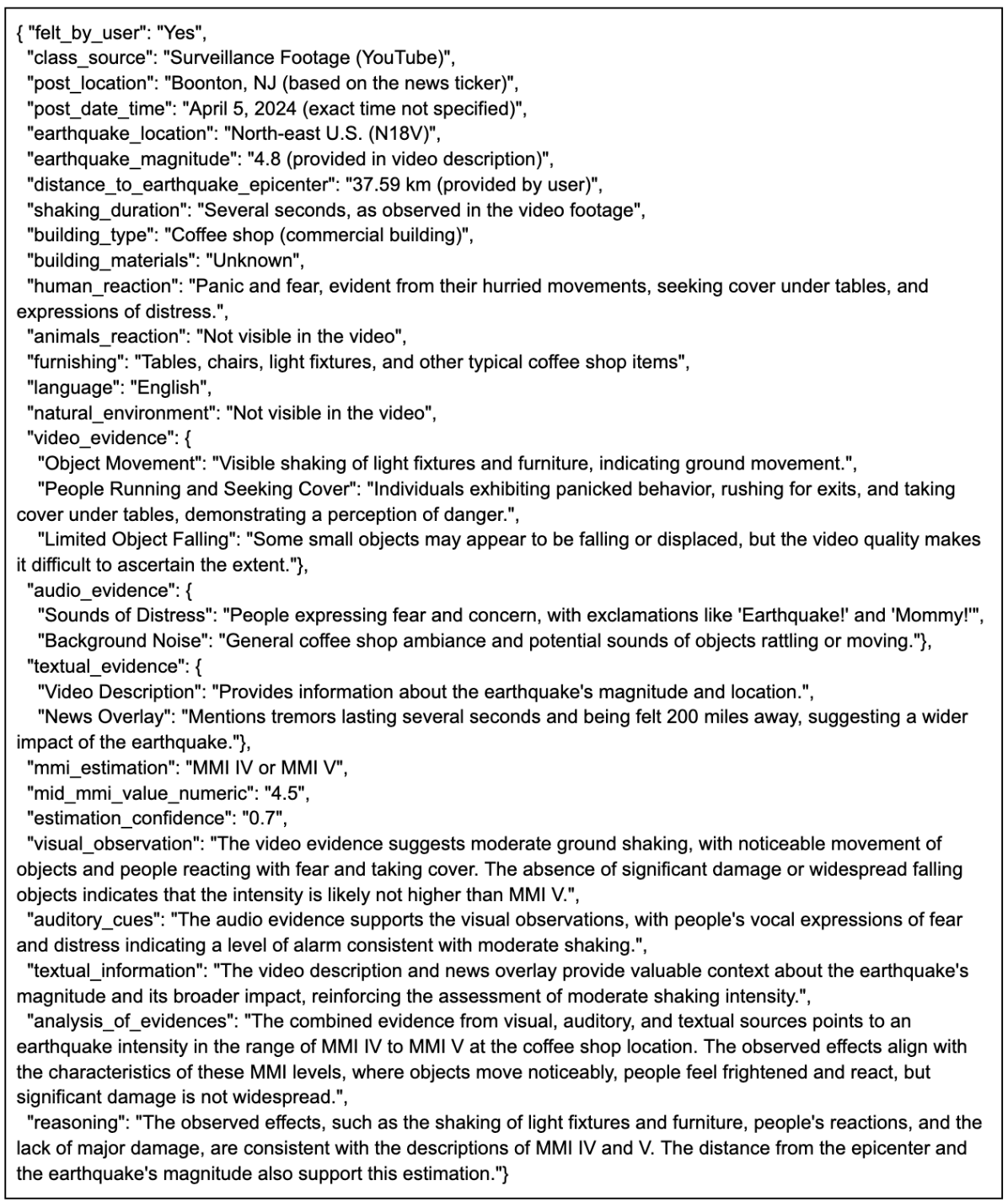}
    \caption{showcases a sample of Gemini's output for a social media post. This post documents the ground shaking experienced in Boonton, NJ, located approximately 37.6 km from the epicenter of the M4.8 earthquake that occurred on April 5, 2024, in Tewksbury, New Jersey, USA.}
    \label{fig:output_sample}
\end{figure}

We used Gemini 1.5 Pro\cite{reid2024gemini}  model and few-shot prompting (see the method section for more details), for processing and analyzing the collected data. Figure \ref{fig:output_sample} showcases an example of Gemini's output, processing and analyzing the contents of a social media post featuring a CCTV video during the 2024 M4.8 earthquake in New Jersey. The model effectively extracts key information such as the post's location, earthquake magnitude, shaking duration, building type, and reactions of humans and animals. It then summarizes and analyzes visual, auditory, and textual cues (e.g., the sound of objects rattling) to estimate the shaking intensity on the MMI scale, providing a rationale for its estimation. The ability to perform cross-modal analysis and acknowledge limitations like incomplete views or unknown epicentral distances is noteworthy. However, it's crucial to recognize that LLMs like Gemini can generate hallucinations, leading to variations in estimated MMI values across different inferences. To address this and obtain more consistent results with uncertainty quantification, we calculate the mean, median, and standard deviation of the estimated MMI values across ten inferences on each sample. Gemini’s outputs for each post, along with the corresponding input data, are available in the supplementary materials of this paper.

To validate our findings, we compared them with independently determined intensities from the "Did You Feel It?" (DYFI) dataset \cite{atkinson2007did, wald2011usgs, quitoriano2020usgs}. This valuable USGS resource collects post-earthquake reports (through online questionnaires) from individuals who experienced the event, providing insights into ground shaking intensity and geographic distribution. Participants share observations through the DYFI website or app, contributing to scientific understanding of earthquake effects. Despite inherent variability and uncertainties, DYFI data serves as a crucial "ground truth" reference for validating Gemini's estimates. Figure 3 illustrates the mean MMI values estimated by Gemini (with ±1 standard deviation) for two well-documented earthquakes (New Jersey and Oklahoma) overlaid with USGS DYFI data, seismogram-based MMI estimations, and the region's expected attenuation model. Gemini's MMI estimates align with the expected intensity range derived from ground motion prediction models, felt reports, and instrumental measurements at comparable distances, supporting the validity of our approach. The highest computed uncertainties in estimated MMI valused is related to a CCTV footage recorded in High Bridge, NJ, ~11.92 km from the epicenter of M4.8 Tewksbury earthquake. This video, with a limited view to the sky and a few trees in an open area, offers limited evidence of shaking intensity such as the tree swaying and lacks other strong audio and textual supports for a more deterministic intensity estimation. Furthermore, Figure \ref{fig:hists} demonstrates that Gemini's mean MMI values for these earthquakes fall within a similar intensity-distance distribution as the DYFI data. Both events, classified as moderate with reported intensities primarily between III and IV, exhibit clustering around population centers like New York City and Tulsa.

The low range MMIs (i.e III to IV) often form the majority of felt reports collected for the moderate earthquakes. To assess the model's ability to generalize to other regions and perceive higher intensity levels, we analyzed data from larger events in diverse locations. However, due to a limited number of available posts, we restricted our analysis to comparing Gemini's estimates with the distributions of DYFI data reported for the city where most of our post data originated (Figure \ref{fig:all_events}). The results demonstrate that Gemini's estimated MMIs generally fall within the interquartile ranges of DYFI reports. Notably, Gemini's intensity estimates for major events (M7.4, 7.5, and 9.1) are consistently and significantly higher than those for moderate events (M4.6, 4.8, and 5.1). The estimated average MMI of 8.4 for the Tohoku earthquake suggests that intensities derived from videos might not exhibit the saturation limitations of macroseismic intensities \cite{bossupotential}. However, further data and case studies are necessary to validate this hypothesis. The most significant discrepancy between our estimates and DYFI data is observed for the M7.4 Taiwan event. This could potentially be attributed to high uncertainties in estimating content source locations and the sparsity of DYFI data for the region.

\begin{figure}[ht!]
    \centering
    \includegraphics[width=0.95\columnwidth]{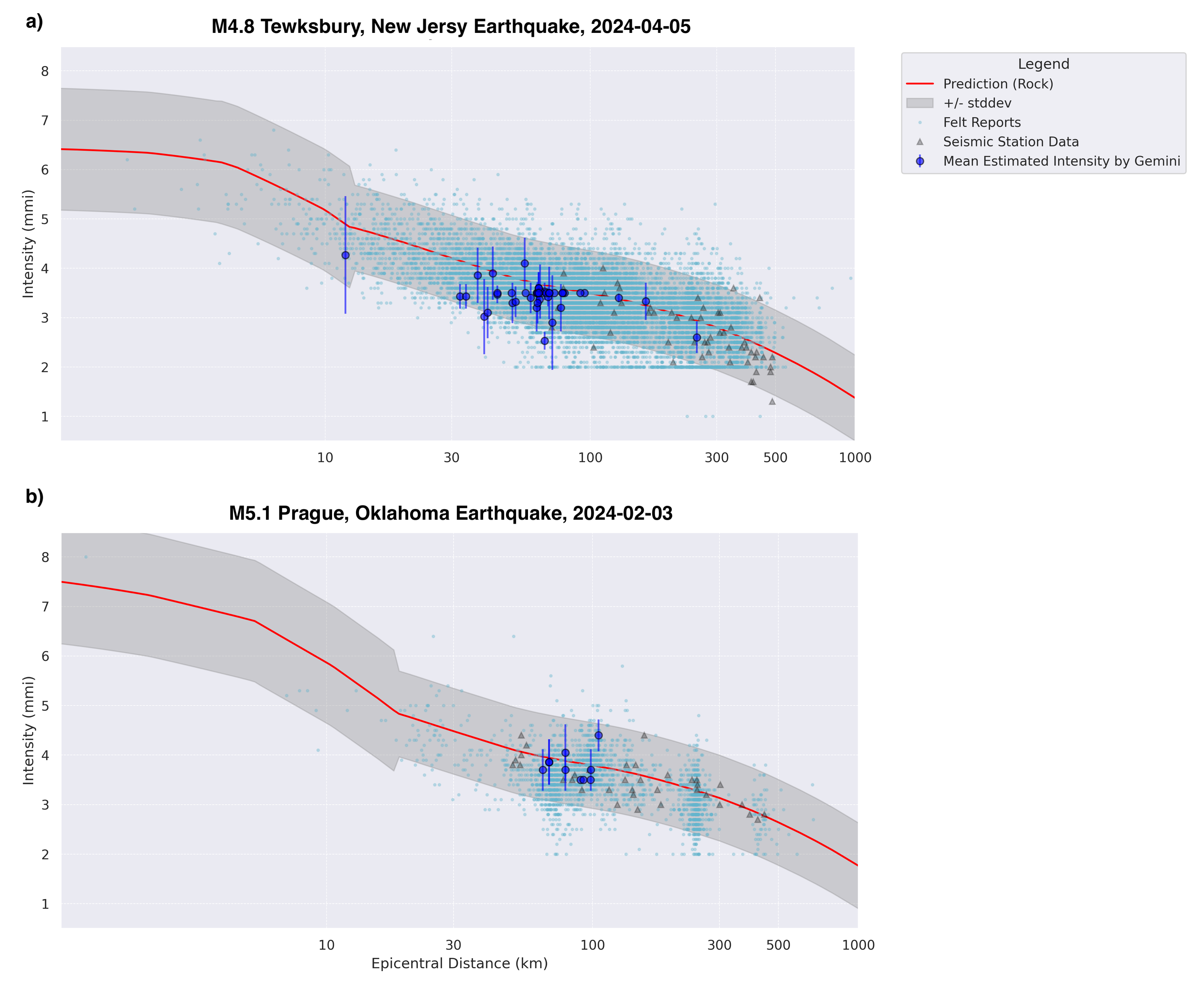}
    \caption{presents a comparison of estimated Modified Mercalli Intensities (MMIs) from the Gemini model (dark blue circles with error bars) against several sources of observed data. These include: (1) instrumentally-derived MMIs computed from peak ground acceleration (PGA) recorded by seismic stations; (2) "Did You Feel It?" (DYFI) macroseismic data collected by the USGS; and (3) the expected ground motion attenuation model for rock sites on the East Coast, along with its ±1 standard deviation range. Panels (a) and (b) display the results for the New Jersey and Oklahoma earthquakes, respectively. Note that the distance scale on the x-axis is logarithmic.}
    \label{fig:regression_plot}
\end{figure}

\begin{figure}[ht!]
    \centering
    \includegraphics[width=0.95\columnwidth]{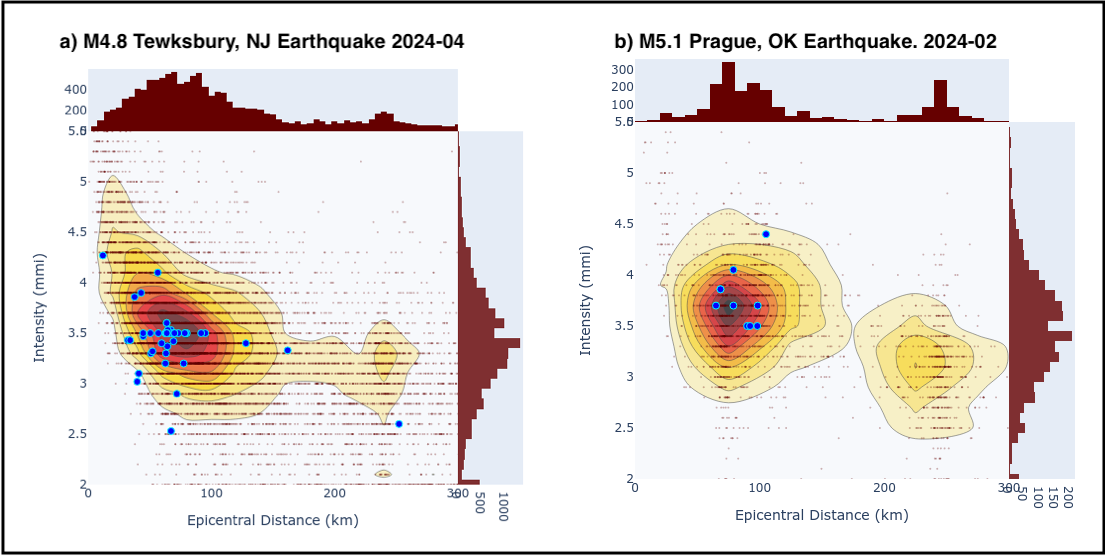}
    \caption{Histograms illustrating the distribution of earthquake intensity with respect to epicentral distance for the New Jersey (a) and Oklahoma (b) earthquakes, utilizing USGS DYFI data. Each panel presents a 2D histogram, while its margins display 1D histograms with counts on the axes. The blue circles represent the estimated mean Modified Mercalli Intensity (MMI) values for individual social media posts, as determined by the Gemini model.}
    \label{fig:hists}
\end{figure}

\begin{figure}[ht!]
    \centering
    \includegraphics[width=0.95\columnwidth]{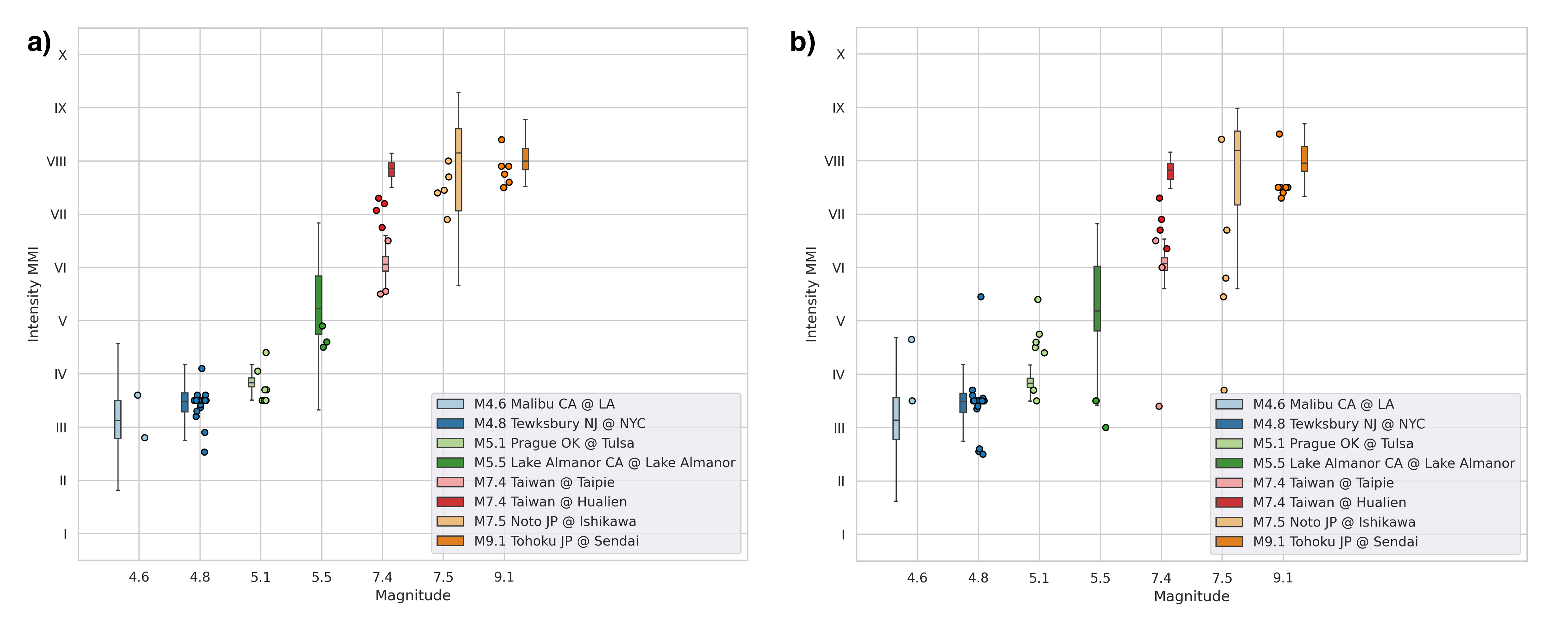}
    \caption{Comparison of Gemini’s estimate with (a) and without (b) epicentral distance and earthquake magnitude in the prompt. Each boxplot with whiskers illustrates the distribution, quartiles, and outliers of reported Did You Feel It? (DYFI) data for individual earthquakes within specific zip codes of a city. Circle markers represent the estimated mean MMI values derived from each social media post, within the same city,  analyzed by the Gemini model for each event.}
    \label{fig:all_events}
\end{figure}

\begin{figure}[ht!]
    \centering
    \includegraphics[width=0.95\columnwidth]{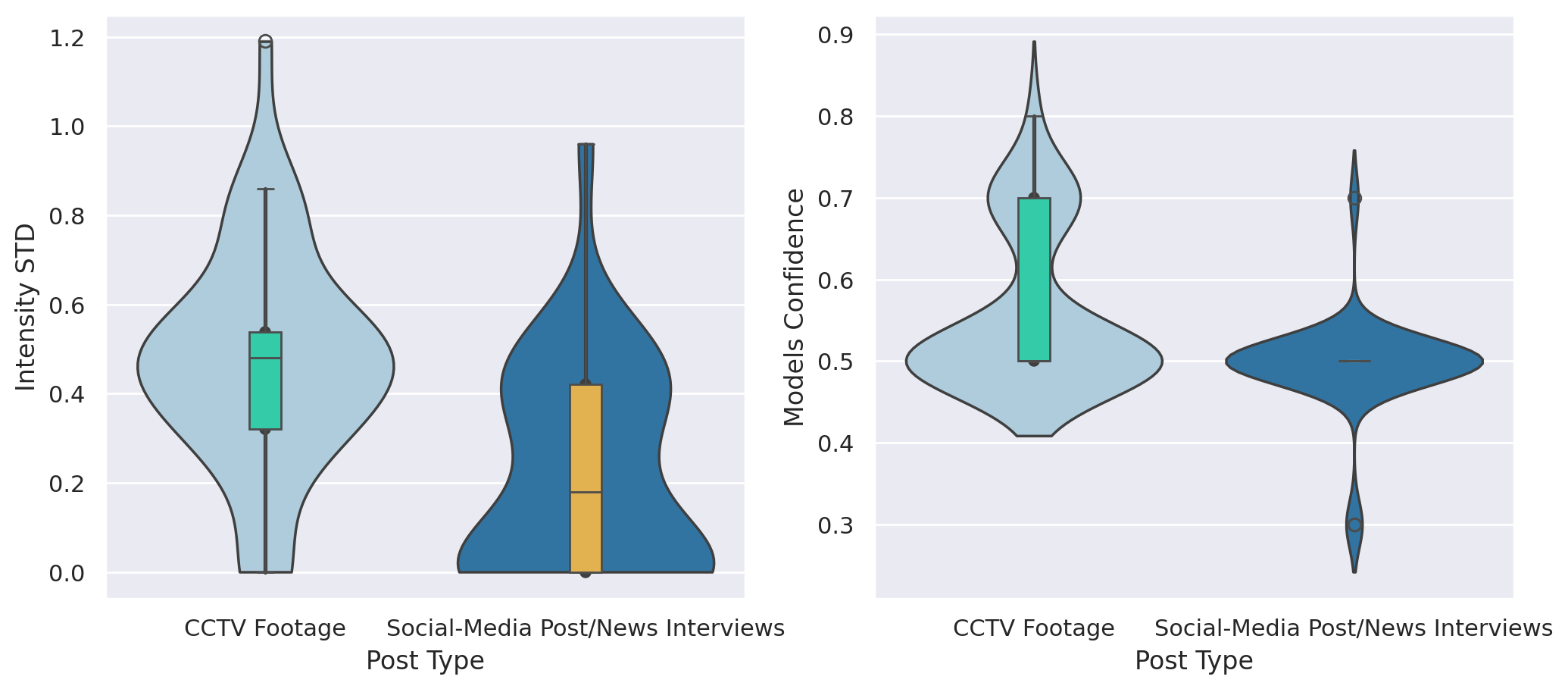}
    \caption{comparison of model uncertainties (a) (variations in Gemini’s estimates across multiple inferences) and the model confidence (provided by Gemini directly) (b)  in intensity estimation from CCTV footage and social-media posts/news interviews.}
    \label{fig:violin_plot}
\end{figure}

\begin{table*}[!htbp]
\caption{{List of earthquakes and the number of posts collected and analyzed for each event in this study.} }
\label{event_table}
\def\arraystretch{1}
\ignorespaces 
\centering 
\begin{tabulary}{\linewidth}{LLLLLLL}
\hline \textit{Earthquake Location} & \textit{arthquake Date TIme (UTC)} & \textit{Magnitude Mw} & \textit{Number of Posts}\\
\hline 
Tohoku, Japan & 2011-03-11 05:46:24 & 9.1 & 7 \\
Lake Almanor, California, USA & 2023-05-11 23:19:41 & 5.5 & 3 \\
Noto Peninsula, Japan & 2024-01-01 07:10:09 & 7.5 & 8 \\
Prague, Oklahoma, USA & 2024-02-03 05:24:28 & 5.1 & 10 \\
Malibu, California, USA &2024-02-09 21:47:27 & 4.6 & 2 \\
Hualien City, Taiwan & 2024-04-02 23:58:12 & 7.4 & 7 \\
Tewksbury, New Jersey, USA & 2024-04-05 14:23:20 & 4.8 & 45 \\
\hline 
\end{tabulary}\par 
\end{table*}

Figure \ref{fig:violin_plot} presents a comparison of estimation uncertainty (through bootstrapping) in Gemini's results for CCTV footage versus social media posts and news interviews. Our estimated model uncertainty indicates a higher variability in model’s estimates derived from CCTV footage relative to those derived from social-media posts (Figure \ref{fig:violin_plot}a). In contrast, Figure \ref{fig:violin_plot}b suggests that the model often rates its estimates from CCTV footage with relatively higher confidence values.  

One possible explanation for the model's higher confidence in CCTV footage could be the richer set of information typically available in videos, compared to the often limited evidence found in social media posts, such as short tweets. However, this could result in a higher variability in the model’s estimates as well, potentially, due to a wider range of observations/evidence available for the analysis and reasoning. We did not observe a strong correlation between the estimated uncertainties and the confidence values provided by Gemini (Supplementary Figure 1). Interestingly, the narrative descriptions of shaking experiences shared by users in social media posts and interviews, although less direct than the captured shaking moments in CCTV footage and reducing the confidence of the model on its estimate, appear to constrain the model's estimates, resulting in lower overall uncertainties comp;ared to estimates from CCTV footages. However, the less deterministic nature of descriptive evidence leads to a wider range of uncertainties compared to the more visually explicit evidence in videos. Factors such as limited views, low-quality audio, and the lack of informative contextual information in some CCTV footage could contribute to the higher uncertainties observed in MMI estimations for this data type. Further investigation is warranted to better understand the observed differences in uncertainty between data types and to explore methods for improving the model's performance and confidence across all sources of information. This could involve analyzing the specific types of descriptive evidence that contribute to lower uncertainties, as well as developing techniques to extract more contextual information from CCTV footage.

\section*{Discussion }

\textbf{How Does Gemini Estimate MMI? }

\begin{figure}[ht!]
    \centering
    \includegraphics[width=0.95\columnwidth]{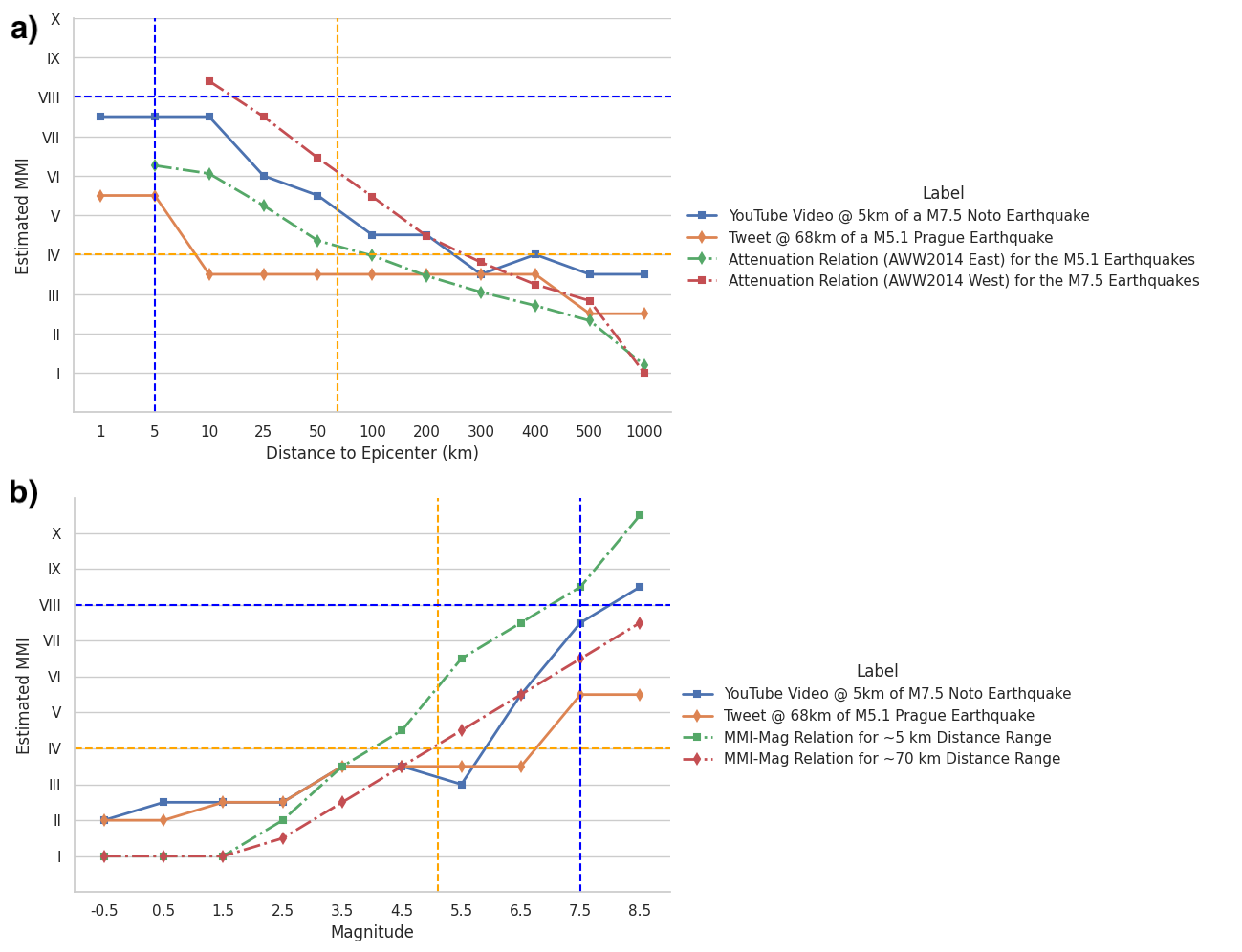}
    \caption{Does in addition to the reading and listening Gemini knows physics? Estimated MMI values by Gemini for two sample inputs (a CCTV video recorded at 5km of M7.5 Noto earthquake and a tweet ~68 km of M5.1 Prague earthquake) but varying provided information on epicentral distance (top) and earthquake magnitude (bottom) in the prompt. The true distance, magnitude, and MMI for each event is depicted by vertical and horizontal lines respectively. The dotted dashed lines are the expected MMI values as a function of distance and magnitude from known empirical relationships.}
    \label{fig:mag_dist_variaiton}
\end{figure}

Our results suggest that Gemini can estimate ground motion intensity from social media posts and CCTV videos with a comparable variability to the estimates based on conventional felt reports and instrumental measurements. This capability stems from Gemini's advanced understanding of language, images, video, and audio. While the model effectively retains detailed and relevant information from the input data in most cases, its capabilities, similar to humans, are not without limitations and can be prone to errors due to factors such as low-quality inputs, limited evidence, and the presence of noise. For example, misinterpretations of visual cues or ambiguous language in social media posts can lead to inaccurate estimations. Beyond its information extraction and understanding abilities, Gemini appears to utilize additional sources of knowledge about general magnitude-distance-intensity relationships of earthquakes, presumably acquired during its training, in its reasoning and decision-making processes. Visual inspection of the model's outputs reveals frequent references to epicentral distance, earthquake magnitude, and post comments in the reasoning section, suggesting that Gemini actively incorporates this contextual information, whether provided directly in the input data or through the prompt, during its analysis. To further investigate the influence of provided context on the output results, we conducted two experiments. In the first experiment, we selected two sample inputs: a CCTV video recorded approximately 5 km from the M7.5 Noto earthquake and a tweet from about 68 km away from the Prague earthquake. In the first experiment, multiple queries were performed on each input sample, varying only the value of the epicentral distance in the prompt while keeping the magnitude fixed at its true value. For both cases, Gemini's estimated MMI value systematically decreased as larger epicentral distances were used in the prompt, aligning with established principles of earthquake physics and empirical relationships in seismology (Figure \ref{fig:mag_dist_variaiton}, top). Similar observations were made when varying the magnitude, the second experiment, while keeping the distance fixed (Figure \ref{fig:mag_dist_variaiton}, bottom). These results indicate that Gemini might have a knowledge of relationships between earthquakes magnitude, distance to the epicenter, and ground shaking intensity and actively use it in its predictions. However, our observations, from performing a similar test to multiple other examples, suggest that Gemini uses this knowledge in conjunction with the evidence it extracts from the provided data. This is why its upper and lower estimation bounds, i.e. MMI 8.5 and 2 for Video example and MMI 5.5 and 2 for the Tweet example respectively, remains within the range of viable intensity levels based on the observations and evidence within the input data. As an example, even when the prompt states that provided Tweet was posted by a person 1000 km away from a M 0.5 earthquake, a case where all the empirical relationships predict MMI I or not felt intensity, Gemini still outputs the minimum felt intensity level (MMI II) as it can not ignore the provided evidence by the user who clearly mentioned feeling of an earthquake.  

\begin{figure}[ht!]
    \centering
    \includegraphics[width=0.95\columnwidth]{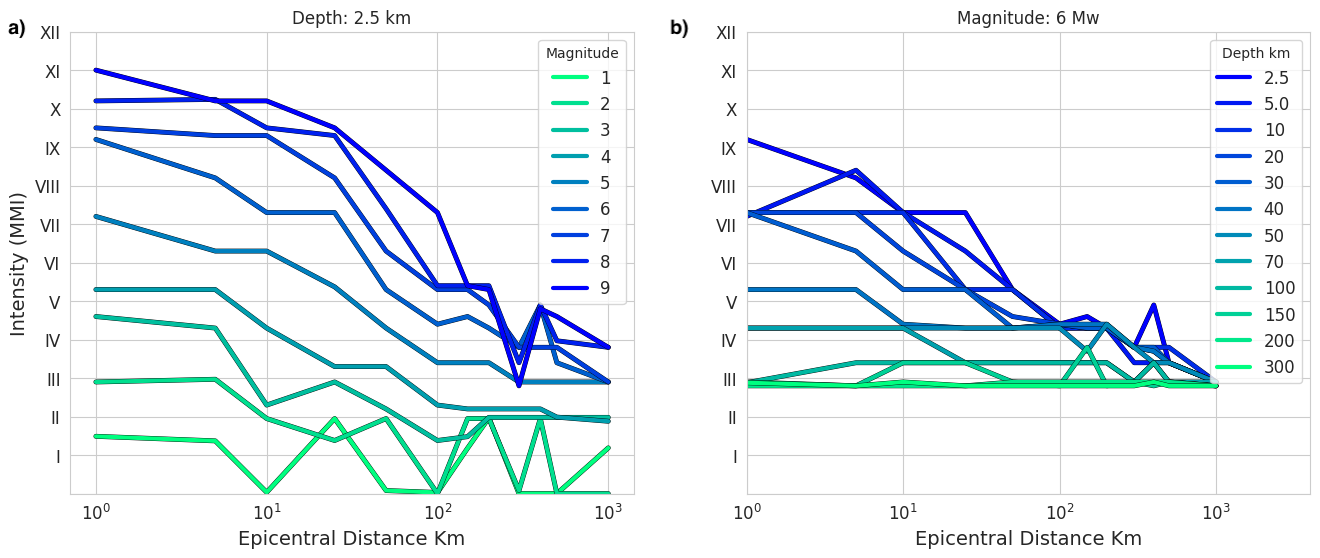}
    \caption{Gemini’s estimates of MMI level at different epicentral distance ranges for varying earthquake magnitudes (left) and depths (right).}
    \label{fig:gemini_relation}
\end{figure}

These findings raise intriguing questions about the extent of Gemini's knowledge and reasoning capabilities. To further investigate this, we simply asked Gemini for its reference and if it uses any particular ground motion prediction equation. Here is its response: \textit{"I did not use a specific ground motion prediction equation (GMPE) to arrive at the MMI estimates. My response was based on a simplified understanding of the general relationship between magnitude, distance, and MMI intensity. This approach does not involve the complexity of GMPEs, which consider various factors like fault type, regional geology, and site conditions to predict ground motion parameters."} To get a feeling for this “simplified understanding of the general relationship” that Gemini has learned about earthquakes, we performed a large number of queries asking for a general  estimate of MMI level at different epicentral distance, earthquake magnitude, and earthquake depth ranges without providing any input data, exemplar, and additional contextual information (Figure \ref{fig:gemini_relation}). The results clearly demonstrate an understanding of the relationship between earthquake characteristics and shaking intensity. It is important to consider that this knowledge likely stems from the vast amount of data it was trained on, rather than an inherent understanding of physics principles. However, more studies, perhaps on more well defined problems with less uncertain ground truths and more available data, are needed to further investigate the extent of Gemini’s general understanding of the physical world and its phenomena.  Nevertheless, the ability to incorporate contextual information and generate results consistent with established scientific knowledge highlights the potential of LLMs like Gemini in augmenting our understanding of complex physical phenomena such as earthquakes.

\textbf{Potential Applications.}

The results of our study indicate LLMs hold immense potential for deriving macroseismic intensity values from crowdsourced eyewitness accounts. Traditionally, macroseismic data collection relied on voluntary participation in structured surveys like "Did You Feel It?". However, flexible approaches using multi-modal LLMs like Gemini as presented here offer potential for significantly scaling up data collection. We can overcome language barriers and the limitations of structured questionnaires by extracting subtle, yet crucial information directly from multimodal data sources. This allows for rapid and flexible approaches that integrate visual, auditory, and textual evidence into the analysis process. The ability to generate scientific measures like earthquake shaking intensity directly from unstructured data such as CCTV footage and social media posts opens exciting new avenues in citizen seismology\cite{bossu2011citizen}) and deep-learning seismology\cite{mousavi2022deep}. This makes collecting valuable scientific data from unconventional resources such as social media posts, open comments, and geo-located pictures and videos more feasible, enriching our understanding of seismic events and their impact. This can be a valuable source of information to constrain and improve these estimates of often uncertain damage scenario estimates solely based on earthquake magnitude and location, particularly in sparsely-instrumented regions. LLMs offer efficient and effective methods for real-time processing and analysis of crowdsourced data during naturald disasters. This can contribute to a better understanding of their impact on communities by providing real-time crisis mapping and rapid situation awareness\cite{middleton2013real}. The extracted intensities (along with improved geolocation data) can be used for various purposes such as: improving the alert models in EEW systems, rapid impact assessment\cite{vieweg2010microblogging, kryvasheyeu2016rapid, bossupotential}, rapid determination of fault geometry\cite{bose2021near}, reducing uncertainties in ShakeMaps\cite{quitoriano2022combining}, improving human loss estimations (e.g., USGS PAGER system), distinguishing high-impact from low-impact events\cite{lilienkamp2023utilization}, providing situational awareness\cite{yin2012using}, protective action decision-making\cite{zhang2024modeling}, and assessing information credibility\cite{castillo2013predicting}.

\textbf{Current Challenges and Potential Solutions.}

Currently, a major challenge/limitation lies in the lack of reliable (and precise enough) location information which could enable a more rigorous testing and evaluation against ground truths. Often, CCTV footage of earthquake shaking goes viral quickly, making it difficult to identify the original source to verify the location of the observation. To fully leverage such systems for crowdsourcing earthquake information, supplementary geoparsing or geotagging techniques can be employed\cite{middleton2018location, huang2019hierarchical }. Similarly, the use of official API might offer a more precise location estimation in some cases. Beside location estimation, to improve the intensity estimations, more sophisticated prompt engineering techniques and a variety of exemplars can be employed. Furthermore, the LLM models can be fine tuned using labeled datasets like DYFI or EMSC datasets. Additionally, incorporating supplementary seismological data, such as recorded ground acceleration, earthquake depths, and historical earthquake evidence at that location into the analysis process might lead to further improvements\cite{burks2014rapid}. This is where the multi-modality of LLMs like Gemini becomes particularly useful. Finally, multi-LLM based intelligent agents could be developed to automate the entire process\cite{boiko2023emergent}.

\textbf{Limitations and Potential Risks. }

While LLMs show promise for earthquake and natural disaster research, offering new avenues for autonomous scientific inquiry, it's crucial to acknowledge their limitations and potential risks\cite{bommasani2021opportunities}. 

Prompt Engineering and its Effects. Like all LLMs, Gemini's output is highly sensitive to the input prompt. Even minor changes can lead to different MMI estimations. For example, adopting different personas in the prompt, such as\textit{"earthquake engineer"} or \textit{"disaster management professional"} can yield to different sets of outputs each tailored to specific professional needs and interests. The high variability of inference outputs and inconsistency of estimates in some cases pose the other main issue. Several techniques can improve the reliability and consistency of LLM outputs. These include calibrating output probabilities\cite{zhao2021calibrate}, using a noisy channel \cite{min2021noisy}, augmenting few-shot examples with intermediate steps \cite{reynolds2021prompt}, or employing the Graph of Thoughts (GoT) technique (Besta et al., 2024) for multi-step reasoning. Furthermore, mining and paraphrasing methods can automatically augment prompt sets \cite{jiang2020can}. Deeper exploration of successful prompting strategies \cite{xie2021explanation} may reveal how to elicit emergent abilities of the models. However, understanding why models work often lags behind the development and popularization of techniques like few-shot prompting. Best practices for prompting are also likely to evolve as more powerful models emerge.

Summation Effect. Gemini 1.5 Pro boasts a remarkable ability to understand information within a long context, allowing it to process lengthy videos containing multiple reports. We investigated the impact of this capability on model output by analyzing a news report featuring interviews with three individuals who experienced the April 2024 M4.8 earthquake in New Jersey. The report was processed in two ways: first, as a single video containing all three interviews, and second, as three separate videos, each trimmed to include only one interview. The estimated MMI for the long video encompassing all three interviews was 4.5, which matched the highest MMI estimate obtained from the individual interview segments (MMI 3.5, 3.5, and 4.5, respectively). When the original video was shortened to include only the first two interviews, the estimated MMI dropped to 3.5. This suggests that the content of the last interview influenced the overall MMI estimation.

Multilingual Capabilities and Limitations. A significant advantage of Gemini is its ability to process content in multiple languages directly. Our dataset incorporates posts and dialogues written in Chinese and Japanese, which the model successfully comprehended and analyzed. However, it's important to note that encountering unsupported languages can lead to inaccurate location estimations (for both the post and the earthquake event) as the model is still able to distinguish the language and wrongly associated with the earthquake vehicle not fully comprehending its content. An interesting observation was that even in such a case, the estimated MMI value would still remain unaffected as it relies on evidence the model can extract and confidently relate to the earthquake shaking.

Irrelevant Information and Estimation Accuracy. Across several experiments, we examined the potential impact of extraneous information, such as advertisements or user comments, on the accuracy of intensity estimations. Our findings revealed no noticeable difference in model performance between posts containing such distractions and those without. This suggests that the models effectively filter irrelevant content, focusing primarily on the pertinent information for estimation. Consequently, the presence of extraneous elements appears to have negligible influence on the models' accuracy.

\begin{methods}

GenAI and LLMs have revolutionized the field of natural language processing\cite{wei2022emergent}. Models such as BERT \cite{devlin2018bert}, BART\cite{lewis2019bart}, and GPT-4\cite{achiam2023gpt} have demonstrated remarkable capabilities in generating coherent and contextually relevant text, translating languages, summarizing documents, and performing various other language-related tasks, often surpassing human performance in benchmarks. In this study, we utilize the Gemini 1.5 Pro\cite{reid2024gemini} model, from Gemini (Gemini Team 2023) family, to process and analyze the collected data.

Gemini (Gemini Team 2023), a state-of-the-art large language model developed by Google, is a suite of generative AI models designed to interpret and respond to user inputs using natural language processing (NLP). What sets Gemini apart is its multi-modality, meaning it can reason across different input data types, including text, audio, images, and video, making it particularly well-suited for analyzing the diverse social media content in our dataset. We chose Gemini 1.5 Pro \cite{reid2024gemini} for this study due to its unique combination of multi-modality processing and long-context understanding. Gemini 1.5 Pro delivers a breakthrough in long-context understanding, with the ability to process up to two million tokens consistently, achieving the longest context window of any large-scale foundation model to date. This expanded context window allows for more comprehensive processing of information, leading to more consistent, relevant, and useful results. These features are particularly valuable for our research, as they allow us to effectively extract insights from the full range of information present in the data and for developing a more nuanced understanding of earthquake impacts.

Gemini, like many large language models, operates through prompting. This involves providing the pre-trained model with a natural language instruction or "prompt" that guides its response generation without requiring further training or parameter updates. The effectiveness of a prompt is crucial in steering the model towards the desired output. A well-crafted prompt should exhibit clarity, conciseness, and sufficient context, enabling the model to accurately comprehend and execute the task. Key elements include explicit task specification, relevant background information, natural language phrasing, illustrative examples, and consistent coherence throughout the prompt. By meticulously designing prompts, users can fine-tune the model's behavior and leverage its capabilities to produce precise and pertinent outputs tailored to specific needs and objectives.

Several prompting techniques exist, including zero-shot, one-shot, few-shot, and multi-shot prompting\cite{brown2020language, kojima2022large}. Zero-shot prompting, also known as direct prompting, provides the model with only instructions and no examples. This approach is well-suited for creative tasks and benchmarking. One-shot prompting offers the model a single concise and descriptive example to guide its output. Few-shot and multi-shot prompting, on the other hand, provide multiple examples, proving more effective for complex tasks requiring pattern replication or specific output structures that are difficult to describe explicitly. Our experiments demonstrate that Gemini can estimate ground shaking intensity based on the content of a social media post even through a simple zero-shot prompt such as: \textit{"Use the video, audio, and text in this social media post shared by a person who felt an earthquake to estimate the intensity of ground shaking at its location in the Modified Mercalli Intensity (MMI) Scale."} This capability stems from the inclusion of MMI scale definitions and information within its training data. However, for improved accuracy and consistency in these estimations, more sophisticated prompt engineering techniques are necessary\cite{wang2023plan}.

The few-shot prompt employed in this study, detailed in Table S1, comprises four key components: background information (persona and context), instruction or query, desired output format, and exemplar. The exemplar serves as a template, outlining the structure of the expected output. This structure includes: 1) summarizing pertinent information extracted from the input data (e.g., post time and location, potential earthquake association, shaking duration, building type, people's reactions); 2) organizing collected visual, auditory, and textual observations and evidence; 3) providing an estimated MMI value; 4) explaining the reasoning behind the estimation process; and 5) acknowledging limitations that may impact the model's estimate. To facilitate complex reasoning, the prompt utilizes the chain-of-thought (CoT) prompting technique\cite{cobbe2021training, suzgun2022challenging}. CoT prompting encourages large language models (LLMs) to break down problems into intermediate steps, mimicking human thought processes. By presenting the LLM with examples that explicitly demonstrate reasoning steps, it is encouraged to follow suit, leading to more accurate and transparent results.

To enhance the reliability and stability of responses, our pipeline incorporates contextual information such as the distance to the earthquake epicenter and the earthquake's magnitude (Figure \ref{fig:pipeline}. Initially, we extract the post's location (city, state, country) from the input video/image using a one-shot prompt: \textit{“Analyze the provided image/video and extract any information that indicates the location of the user or post. Identify the specific location name, including city, state (if applicable), and country. Limit your response only to the extracted location. Example output: Imperial, CA”}. Following this, we automatically calculate the distance from the earthquake epicenter. Both the earthquake magnitude and the computed epicentral distance are then integrated into the main prompt to query the full intensity analysis.

\begin{figure}[ht!]
    \centering
    \includegraphics[width=0.95\columnwidth]{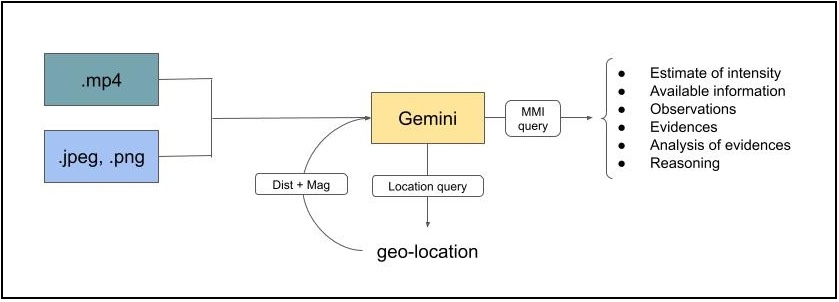}
    \caption{illustrates the processing workflow employed in this study. Initially, images and videos are input into the model. A basic query is then executed to extract available evidence pertaining to the location of observation. This evidence can manifest in textual, visual, or auditory formats. The extracted location information is subsequently utilized to calculate the distance (great circle in kilometer) to the earthquake's epicenter. This distance, along with the earthquake's magnitude, serves as input for the next prompting stage in which a comprehensive intensity analysis is conducted.}
    \label{fig:pipeline}
\end{figure}

We conducted our analysis using Vertex AI, a Google Cloud machine learning (ML) platform that facilitates the training, deployment, and customization of AI applications and ML models. Our pipeline leveraged the Gemini 1.5 Pro model with specific parameters to ensure consistency and control over the generated outputs. We set the temperature value to 0.5, which promotes more deterministic and probability-driven predictions, favoring the most likely tokens and reducing randomness. Additionally, we maintained the default output token limit of 8,192, equivalent to approximately 32,768 characters, to accommodate the desired output length. Finally, we employed a top-p value of 0.95, which dictates the selection of tokens based on their cumulative probability distribution, ensuring diversity while maintaining focus on the most probable options.

\textbf{Data Availability}

The data used in this study were collected from public accounts on YouTube, Twitter, and TikTok. The dataset and Gemini’s results  can be found here: (The linik will be provided). Maximum intensity “Did You Feel It?” data for each event were collected from USGS website, https://earthquake.usgs.gov/earthquakes/eventpage/us7000ma74/dyfi/responses (last accessed May 2024

\textbf{Code Availability}

Our source code and model are available at \BreakURLText{https://github.com/smousavi05/}.

\clearpage
\section*{References}
\bibliographystyle{naturemag}
\bibliography{bibliography.bib}

\textbf{Acknowledgements}
We are thankful to Roy Want and Mohammed Khider for insightful comments and suggestions. Some of the python codes for plotting the results of this study have been generated by Gemini. Figures in this paper were generated using the Seaborn (Waskom, 2021) and Matplotlib (Hunter, 2007).

\textbf{Authors' contributions}

 S.M.M. designed the study, compiled the dataset, implemented the method, performed the analyses, and wrote the initial draft of the manuscript. All authors discussed extensively the results and contributed to the final version of the manuscript.\end{methods}
    
\textbf{Competing interests} 

The authors declare no competing interests.

\end{document}